\documentclass[conference,a4paper]{IEEEtran}
\IEEEoverridecommandlockouts

\usepackage{cite}
\usepackage{amsmath,amssymb,amsfonts}
\usepackage{algorithmic}
\usepackage{graphicx}
\usepackage{textcomp}
\usepackage{xcolor}

\usepackage{tikz}
\usetikzlibrary{positioning}

\newcommand{\url}[1]{#1}

\def\BibTeX{{\rm B\kern-.05em{\sc i\kern-.025em b}\kern-.08em
    T\kern-.1667em\lower.7ex\hbox{E}\kern-.125emX}}

\usepackage{color}
\definecolor{gray}{rgb}{0.5,0.5,0.5}
\usepackage[utf8]{inputenc}
\usepackage{units}
\usepackage{url}

\newcommand{\ie}{i.\,e.}

\newcommand{\eg}{e.\,g.}

\begin{document}
\title{TheFragebogen: A Web Browser-based Questionnaire Framework for Scientific Research}

\author{\IEEEauthorblockN{Dennis Guse}
\IEEEauthorblockA{\textit{TU Berlin}\\
dennis.guse@alumni.tu-berlin.de}
\and
\IEEEauthorblockN{Henrique R. Orefice}
\IEEEauthorblockA{\textit{TU Berlin}\\
h.orefice@gmail.com}
\and
\and
\IEEEauthorblockN{Gabriel Reimers}
\IEEEauthorblockA{\textit{TU Berlin}\\
g.reimers@mailbox.org}
\and
\IEEEauthorblockN{Oliver Hohlfeld}
\IEEEauthorblockA{\textit{BTU Cottbus--Senftenberg}\\
oliver.hohlfeld@b-tu.de}
}

\newcommand{\afblock}[1]{\noindent{\textbf{#1}}}
\IEEEoverridecommandlockouts \IEEEpubid{\makebox[\columnwidth]{XXX-X-XXXX-XXXX-0/19/\$31.00 \copyright 2019 IEEE \hfill} \hspace{\columnsep}\makebox[\columnwidth]{ }}
\maketitle

\begin{abstract}
\emph{Quality of Experience}~(QoE) typically involves conducting experiments in which stimuli are presented to participants and their judgments as well as behavioral data are collected.
Nowadays, many experiments require software for the presentation of stimuli and the data collection from participants.
While different software solutions exist, these are not tailored to conduct experiments on QoE.
Moreover, replicating experiments or repeating the same experiment in different settings (\eg, laboratory vs. crowdsourcing) can further increase the software complexity.
TheFragebogen is an open-source, versatile, extendable software framework for the implementation of questionnaires --- especially for research on QoE.
Implemented questionnaires can be presented with a state-of-the-art web~browser to support a broad range of devices while the use of a web~server being optional.
Out-of-the-box, TheFragebogen provides graphical exact scales as well as free-hand input, the ability to collect behavioral data, and playback multimedia content.
\end{abstract}

\begin{IEEEkeywords}
survey software, data collection, experiments
\end{IEEEkeywords}
\begin{tikzpicture}[overlay, remember picture]
\path (current page.north) node (anchor) {};
\node [below=of anchor] {%
2019 Eleventh International Conference on Quality of Multimedia Experience (QoMEX)};
\end{tikzpicture}

\section{Introduction}
In \emph{Quality~of~Experience}~(QoE) research, questionnaires are a fundamental research instrument to gather data in experiments.
In such experiments, participants are presented with a set of stimuli that cover the desired spectrum of degradations, and their perceived quality is captured with questionnaires~\cite{raake_quality_2014}.
In this paper, we present the open-source software framework
\emph{TheFragebogen} that is especially designed (but not limited to) usage in this research domain.
TheFragebogen enables implementing questionnaires to be executed by a web~browser, which allows to support a broad range of devices (even to run multi-device experiments) and simplifies reproducibility.

\afblock{Structure.}
We begin by describing the design of TheFragebogen and its features.
To highlight the benefits of TheFragebogen, we present several experiments it was used in.
Finally, we present lessons learned and give an outlook on future work.

\section{TheFragebogen}\label{sec:thefragebogen}
TheFragebogen has three design principles:
\begin{itemize}
	\item focus on research needs (\eg, scales and multimedia)
	\item extendability \& flexibility (\eg, server and serverless use)
	\item robustness \& reproducibility (\eg, long-term archival)
\end{itemize}
In TheFragebogen, questionnaires are implemented as web~pages using \emph{JavaScript}, \emph{Cascading Style Sheets}~(CSS), and \emph{Hypertext Markup Language}~(HTML).
For portability, it uses standardized web technologies that are available on a broad range of devices.
Further, the text-based nature of these technologies enables using \emph{version-control software}.
While servers can be used \emph{in addition}, its principle design is \emph{serverless}.
Questionnaires can thus be archived, re-used, or shared, and thus hopefully reduce the effort to reproduce experiments.

There exists already a broad range of (commercial) tools for surveys and questionnaires (\eg, \emph{LimeSurvey.org} and \emph{SurveyJS.io}).
However, available frameworks commonly focus on commercial needs rather than research needs (\eg, no/limited support for scales and the ability to capture behavioral data) --- a gap that we aim to close with TheFragebogen.

\begin{table}
    \caption{TheFragebogen technology and links}
    \begin{center}
    	\begin{tabular}{| l | l |}
    	\hline
    	Software license & MIT License \\
    	\hline
    	Programming language & JavaScript (ES5/ES6), CSS3, HTML5 \\
    	\hline
    	\hline

    	Project web page & \url{http://TheFragebogen.de} \\
    	\hline
    	Live demos & \url{http://TheFragebogen.de/demo} \\
    	\hline
    	Software releases & \url{http://TheFragebogen.de/releases} \\
    	\hline
    	Source code repository & \url{http://TheFragebogen.de/code} \\
    	\hline
    	Issue tracker & \url{http://TheFragebogen.de/issues} \\
    	\hline
    	\end{tabular}
    \end{center}
	\vspace{-2.5em}
\end{table}

\subsection{Features to simplify conducting experiments}
We next describe TheFragebogen's core features that address the needs of scientific experiments and in combination differentiate it from existing software solutions.

\afblock{Scales.}
TheFragebogen provides a set of built-in scales using HTML~UI components and also implements several scales that require an exact graphical representation.
At the time of this writing, this includes: the \emph{NASA task load index}~(TLX)~\cite{hart_development_1988}, the \emph{visual analogue scale}~(VAS), and the \emph{continuous rating scale for perceived quality}~\cite[p.\,19]{itu-t_recommendation_p.851_subjective_2003}.
This is especially important as replacing graphical scales with non-similar visualizations can have undesired and potentially unknown effects on the questionees' responses~\cite{funke_webExperiment_2016}.
New graphical scales can be added as \emph{scalable vector graphics}~(SVG).

\afblock{Free-hand input.}
TheFragebogen provides support for free-hand input.
This enables to capture handwritten responses if a digitizer pen on a tablet computer is used.
Using this approach, pen-and-paper questionnaires can be directly recreated without implementing the complete functionality.

\afblock{Behavioral data.}
TheFragebogen allows  capturing behavioral data (\eg, time to answer a question).
This enables, investigating observed anomalies or verify that a participant fulfilled experimental requirements (\eg, in crowdsouring).

\afblock{Multimedia content.}
TheFragebogen provides features for presenting multimedia content (\ie, audio and video playback) while capturing stalling and playback statistics.

\afblock{Remote interaction.}
For interaction with other technical systems, TheFragebogen provides components to send and receive data via WebSockets as well as \emph{Asynchronous JavaScript and XML}~(AJAX).
This enables multi-device experiments (\eg, presenting stimuli and questionnaires on different devices).

\afblock{Data export.}
Collected data, incl. questionees' responses, can be downloaded to the device presenting the questionnaire or uploaded to a web~server.
As format \emph{comma-separated value}~(CSV) was chosen as it is widely used.
Images (\eg, produced by free-hand input) are encoded as \emph{Data URIs}.

\afblock{Extensible design.}
TheFragebogen uses \emph{object-oriented programming}~(OOP) for realizing its components.
This simplifies the development and maintenance process since new components can be added based upon existing ones.

\afblock{Dynamic execution and printability.}
TheFragebogen enables implementing dynamic questionnaires (\ie, subsequent content is adjusted depending on previous answers).
Moreover, questionnaires may be printed to be used as pen-and-paper questionnaires (\eg, as fallback).

\afblock{Respect for privacy.}
TheFragebogen enables gathering data while allowing to collect and store the data without sharing it with a third party (\eg, a cloud provider).
This is especially important considering that the collected data may be sensitive.

\subsection{Lifecycle \& implementation}
A TheFragebogen questionnaire is a \emph{single-page application}.
In this application, the questionnaire is configured using JavaScript (\eg, defining the items and their presentation order).
If desired, CSS can be used to adjust the layout and design.
On execution, TheFragebogen creates the UI components and manages the state of the questionnaire.
Therefore, the use of a web~server is optional.

\begin{figure}
  \centering
  \includegraphics[width=1\columnwidth]{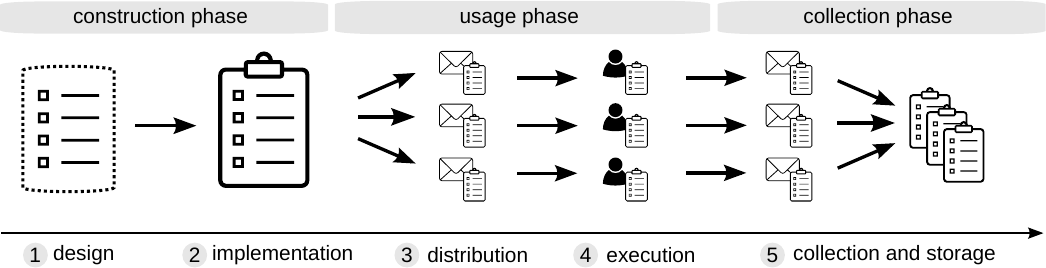}
  \caption{General lifecycle of digital questionnaires.}
  \label{lifecycle}
\vspace{-1em}
\end{figure}

The typical lifecycle of a questionnaire created with TheFragebogen is shown in Figure~\ref{lifecycle}.
It consists of the \emph{construction~phase}, the \emph{usage~phase}, and the \emph{collection~phase}.
In the construction~phase, the questionnaire is designed and prepared for usage.
This also includes the implementation, testing, and, if necessary, setup of required infrastructure.
Tailoring content to subgroups of participants (\eg, randomization) can be implemented using preprocessor software (\eg, Python or shell scripts).
In the usage~phase, the questionnaires are distributed to the questionees.
In the collection~phase, the completed questionnaires are collected and the gathered data is stored for subsequent evaluation.
In these two phases, system failure is problematic as this may lead to data loss.
This includes implementation errors, compatibility issues with web~browser(s), and varying network connectivity.

TheFragebogen follows the \emph{paper metaphor}, \ie, a paper-based questionnaire is actually a sequence of paper sheets, which each consists of one or more questionnaire items.
An item consists, in fact, of a question and a scale.
Sheets are represented in TheFragebogen by so-called \emph{Screens} and items by so-called \emph{QuestionnaireItems}.
A questionnaire consists of one or more \emph{Screens} while only one is shown at a time.
\emph{Screens} encapsulate specific functionalities, such as presenting items, waiting for some time, or exporting the collected data.
A \emph{Screen} is usually presented in full-screen.
The lifecycle of a questionnaire is handled by the \emph{ScreenController}, which organizes the presentation of all \emph{Screens} and data consistency.

\section{Use cases: experiments using TheFragebogen}\label{sec:usecases}
To highlight TheFragebogen's suitability and versatility for use in research, we present conducted experiments investigating different aspects of QoE.
In these, TheFragebogen was used and they also largely influenced its software design.

\subsection{Single device experiments}
Typical experiments are executed on a \emph{single device} only (\eg, tablet or computer).
Here, stimuli are presented on the same device that captures the participant's judgments.

\afblock{Audio stimuli.}
TheFragebogen was used to assess the perceived quality of preprocessed audio stimuli~\cite{guse_duration_2016}.
Here, a tablet computer and a pair of headphones were used.
A tablet computer was chosen as participants could use a digitizer pen to interact with the questionnaire.
TheFragebogen was configured to present one stimulus and subsequently present the desired item(s).
These included category rating scales and text~fields. The later captured a qualitative description of the presented stimuli.
Participants answered these in their own hand writing.
At the end of the experiment, the collected data was downloaded onto the tablet computer.

\afblock{Web browsing.}
Similarly, the perception of the \emph{loading delay} and \emph{load failures} of web pages was investigated~\cite{guse_subjective_2015}.
For this experiment, TheFragebogen was extended to provide UI~components that simulate load~delay and load~failures of images.
This enables to produce these conditions and therefore allows to reproduce the degradations.
The stimuli were presented in a randomized order to every participant.
Randomization was created in the construction~phase using a template questionnaire and a script to generate the randomized questionnaires.
By archiving these questionnaires, the experiment for each participant can be reproduced including the presentation order of stimuli.

\subsection{Experiments using multiple devices}
TheFragebogen was also used in experiments requiring a multi-device setup.
In these, the questionnaire and stimuli are presented on different devices.
This was achieved by using multiple TheFragebogen instances.
This approach was used for the assessment of degraded video stimuli on a mobile device while capturing the responses on a tablet computer~\cite[p.\,42ff.]{guse_multi-episodic_2016}.
Participants used the devices alternatingly, as stimuli and items were presented subsequently.
In this experiment, both devices were connected to a WiFi hotspot and exchanged the experimental progress via WebSockets.
A similar setup was used to investigate the impact of parallel tasks on QoE~\cite{guse_web-qoe_2014}.

TheFragebogen was also used to conduct experiments on conversational speech telephony~\cite[p.\,37ff.]{guse_multi-episodic_2016}.
In these, a pair of participants conducted a series of telephone calls together.
For simulation of the telephony connection incl. degradations, the audio processing software \emph{TheTelephone}\footnote{\url{https://github.com/TheTelephone}}, executed on a separate computer, was used.
The participants' headsets were connected via analogue HiFi~cables and microphone amplifiers to this computer.
Each participant used a tablet computer running one TheFragebogen instance throughout the experiment.
These synchronized their experimental progress incl. presentation of tasks and items as well as informing the audio processing computer about the degradation to be applied.

\subsection{Remote participation}

\afblock{Field experiments.}
TheFragebogen was also used in experiments outside of controlled laboratory settings.
In a 6~day experiment~\cite[p.\,66]{guse_multi-episodic_2016}, questionnaires were distributed using a web~server that could be accessed by the participants with their private computers.
On access by a participant, the correct questionnaire was selected depending on his experimental progress.
To avoid issues due to varying network conditions all multimedia stimuli were preloaded.
After finishing the questionnaire, the collected data was uploaded to the web~server.
On failure, the data was downloaded to the participant's computer with the instruction to send it per email to the researchers.
For this experiment, participants were required to use \emph{Google~Chrome} limiting the testing effort required.

\afblock{Crowdsourcing experiments.}
TheFragebogen was also used in crowdsourcing experiments.
For example, in an experiment to assess the influence of network protocols on the perceived web page loading speed~\cite{zimmermann_qoe_2017}.
To ensure that every participant rated the same stimuli, the captured load processes were presented as videos.
In line with best practices on crowdsourcing~\cite{HossfeldBestPractices}, filtering mechanisms were realized by capturing behavioral data.
For example, to ensure that stimuli were presented completely, focus lost events were captured.
Without modification, TheFragebogen's design enabled to directly repeat the experiment in a laboratory environment to compare the crowdsourcing results with the laboratory results.

\subsection{Lessons learned}
Each of the presented experiments required adding new and refining existing features of TheFragebogen.
While using and extending TheFragebogen, we learned to consider practical aspects to avoid failures.
First of all, technical complexity should be limited if possible and potential issues while running an experiment should be anticipated (\eg, limited network connectivity, varying web~browser).
In addition, it is often beneficial to orchestrate experiments using TheFragebogen.
This avoids human mistakes, for example, if other systems need
 to be re-configured within an experiment.
With regard to long-term use of implemented questionnaires, it is preferable to use established, wide-spread technologies.
Browser-specific functionality or work-in-progess technologies tend to change and then break so far functioning implementations.

\section{Conclusion}
We presented TheFragebogen, an open-source software framework for creating questionnaires.
It is portable, versatile, and extendable as well as out-of-the-box suited for a wide range of use cases in research (\eg, serverless single device questionnaires to crowdsourcing).
TheFragebogen evolved by being used in a number of experiments investigating QoE from different aspects.
By openly releasing it, we aim to ease future experiments that benefit from the practical experience that went into its software design and its feature set.

We hope TheFragebogen may be one aspect to foster reproducibility of experiments.
This could be achieved if researchers openly share their implementations as well as extensions to TheFragebogen.
We believe that a shared software solution can be beneficial for the QoE research community, as it promotes exchange of technical expertise and distributes the effort for extending and maintaining the software.

\section*{Acknowledgements}
Lucas~de~Araujo, Maxim~Spur, and Rodolfo~C.~Dalapicola contributed to TheFragebogen implementation.
Anna~Wunderlich, Falk~R.~Schiffner, Marc~Seebode, Marc~Halbbrügge, Babak~Naderi, and Sebastian~Schuck supported TheFragebogen by using and testing it as well as by requesting new features.
While working on TheFragebogen Henrique~R.~Orefice, Lucas~de~Araujo, and Rodolfo~C.~Dalapicola were funded by the \emph{Ci\^encia sem Fronteiras Alemanha program}. %

\bibliographystyle{IEEEtran}
\bibliography{Bibliography}

\end{document}